\UseRawInputEncoding

\documentclass[reprint, twocolumn, showkeys]{revtex4}

\usepackage{graphicx}
\usepackage{dcolumn}
\usepackage{bm}
\usepackage{color}
\usepackage{amsmath,amsthm}
\usepackage{amssymb,bm}
\usepackage{mathrsfs}
\usepackage{comment}
\usepackage{ulem}

\graphicspath{%
    {converted_graphics/}
    {/}
}
\begin{document}

\title{Bose enhancement of excitation-energy transfer with molecular-exciton-polariton condensates}
\author{Nguyen Thanh Phuc}
\email{nthanhphuc@moleng.kyoto-u.ac.jp}
\affiliation{Department of Molecular Engineering, Graduate School of Engineering, Kyoto University, Kyoto 615-8510, Japan}

\begin{abstract}
Room-temperature Bose--Einstein condensates (BECs) of exciton polaritons have been realized in organic molecular systems owing to the strong light--matter interaction, strong exciton binding energy, and low effective mass of a polaritonic particle. 
These molecular-exciton-polariton BECs have demonstrated their potential in nonlinear optics and optoelectronic applications. 
In this study, we demonstrate that molecular-polariton BECs can be utilized for Bose enhancement of excitation-energy transfer (EET) in a molecular system with an exciton donor coupled to a group of exciton acceptors that are further strongly coupled to a single mode of an optical cavity. 
Similar to the stimulated emission of light in which photons are bosonic particles, a greater rate of EET is observed if the group of acceptors is prepared in the exciton-polariton BEC state than if the acceptors are initially either in their electronic ground states or in a normal excited state with an equal average number of molecular excitations. 
The Bose enhancement also manifests itself as the growth of the EET rate with an increasing number of exciton polaritons in the BEC.
Finally, a permutation-symmetry-based approach to suppress the EET to the huge manifold of dark states in the acceptor group is proposed to facilitate the Bose-enhanced EET to the polariton BEC. 
\end{abstract}

\keywords{Bose enhancement, condensate, excitation energy transfer, molecular polariton}

\maketitle

\textit{Introduction--}Since their first realizations in ultracold atomic gases~\cite{Anderson95, Davis95}, Bose--Einstein condensates (BECs) have been observed in different types of bosonic systems, including semiconductor exciton polaritons~\cite{Kasprzak06}, magnons~\cite{Demokritov06}, photons~\cite{Klaers10}, and plasmons~\cite{Hakala18}. 
Bose--Einstein condensation occurs when the de Broglie wavelength $\lambda_\text{dB}\propto 1/\sqrt{m k_\text{B}T}$ of a particle with mass $m$ and temperature $T$ ($k_\text{B}$ denotes the Boltzmann constant) exceeds the interparticle spacing~\cite{Pethick-book, Pitaevskii-book}. 
Because of the hybridization of light and matter, exciton polaritons inherit the small effective mass of photons confined in an optical cavity.
Moreover, owing to the strong binding energy of Frenkel excitons and large oscillator strength in molecular materials, BECs of molecular exciton polaritons can be observed and studied at room temperature~\cite{Cohen10, Plumhof14, Daskalakis14, Cwik14, Dietrich16, Keeling20}. 
BEC refers to a coherent state in which a single mode is macroscopically occupied, leading to enhanced coherence in both space and time~\cite{Daskalakis15, Betzold20}. 
This coherence in BEC gives rise to various fascinating properties, including superfluidity~\cite{Amo09, Lerario17} and quantized vorticity~\cite{Lagoudakis08, Sanvitto10}. 
Examples of optoelectronic devices based on exciton-polariton condensates include low-threshold lasers and optical switching~\cite{Cohen10, Dietrich16, Sanvitto16, Rajendran19, Wei19}.
Other applications of exciton-polariton BECs are polaritonic simulators of classical spin systems~\cite{Berloff17}. 

On the other hand, polaritons can modify the physical and chemical properties of molecular systems significantly through the strong coupling of electronic or vibrational molecular excitations to an optical cavity~\cite{Ebbesen16, Ribeiro18, Feist18, Hertzog19, Herrera20, Nagarajan21}.
This strong coupling results in various interesting phenomena including the manipulation of chemical landscapes in the excited-state manifold~\cite{Hutchison12, Herrera16, Galego16, Takahashi19}, modification of chemical reactivity by molecular-vibration polaritons~\cite{Thomas16, Hiura18, Thomas19, Lather19, Hirai20, Lather21, Galego19, Angulo19, Phuc20, Li20, Li21, Yang21}, long-range energy transfer~\cite{Feist15, Schachenmayer15}, cavity-enhanced conductivity in organic media~\cite{Orgiu15}, and superreaction~\cite{Phuc21}. 
However, the effects of the already created molecular-exciton-polariton BECs on chemical reactivity have not yet been investigated. 
Notably, there was a recent study on the consequences of a molecular-vibration-polariton BEC, which has not yet been experimentally achieved, on the electron transfer process~\cite{Sivajothi21}. 

In this study, we investigate the excitation-energy transfer (EET) in a molecular system with an exciton donor coupled to a group of exciton acceptors that are further strongly coupled to a single mode of an optical cavity to form molecular-exciton polaritons. 
Through both an analytical investigation and a numerical simulation of the dynamics of the system, we demonstrate that the EET is enhanced by the presence of a molecular-exciton-polariton BEC. 
We first consider the weak-donor-acceptor-coupling limit, where Fermi's golden rule can be applied to obtain the analytic expression for the EET rate, before using the hierarchical equation of motion (HEOM)~\cite{Tanimura06} to numerically investigate the exciton transfer dynamics under a more general condition that goes beyond the perturbative and Markovian limits.
The maximum rate of EET is observed if the group of acceptors is prepared in the molecular-exciton-polariton BEC state, in comparison with the case of the acceptors being initially either in their electronic ground states or in a normal excited state with an equal average number of molecular excitations.  
The Bose enhancement is also justified by the growth of the EET rate with an increasing number of exciton polaritons in the BEC.
By investigating the dependence of EET dynamics on the light--matter coupling strength, it was observed that despite the EET rate being almost independent of the collective Rabi frequency within a short time, a lower EET rate was observed after a long time for a weaker light--matter interaction.
This decrease in the EET rate can be attributed to the effect of decoherence that can be mitigated by a sufficiently strong molecule--cavity coupling through the polaron decoupling effect~\cite{Phuc21}.
Finally, the effect of the dark-state manifold on the EET is discussed.
By utilizing the permutation-symmetry property of these dark states, it can be shown that the EET to the acceptor's dark-state manifold can be suppressed to facilitate the Bose-enhanced EET to the polariton condensate.  

\begin{figure}[tbp] 
  \centering
  \includegraphics[keepaspectratio]{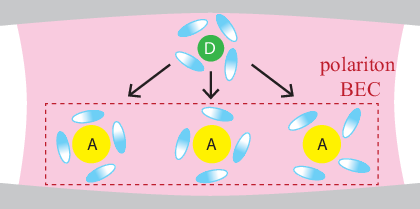}
  \caption{Schematic illustration of EET in a system of an exciton donor (green) coupled to a group of exciton acceptors (yellow) that are further strongly coupled to a single mode of an optical cavity. The collective coupling between the electronic excitations of the acceptors and the cavity field (light magenta) generates molecular-exciton polaritons. These polaritons, which are bosonic particles, can form a BEC in which a single mode is macroscopically occupied. The blue ellipsoids surrounding the molecules represent the environments in which thermal dynamics give rise to energy fluctuations in the molecular system.}
  \label{fig: system}
\end{figure}

\textit{System}--As illustrated in Fig.~\ref{fig: system}, we employed a model molecular system consisting of an exciton donor coupled to a group of $N$ exciton acceptors by the dipole--dipole interaction. 
The acceptors were further coupled to a single mode of an optical cavity. 
The electronic excitation of each molecule is assumed to be accompanied by a change in the configuration of an independent molecular environment, whose normal modes are modeled by a collection of harmonic oscillators.
The total Hamiltonian of the molecular system and the environment is given in the second quantization by $\hat{H}_\text{m}=\hat{H}_\text{D}+\hat{H}_\text{A}+\hat{H}_\text{DA}+\hat{H}_\text{e}$ with the donor Hamiltonian 
\begin{align}
\hat{H}_\text{D}=\left[
\hbar\omega_\text{D}+\sum_\chi g_\chi^\text{D} 
\left(\hat{b}_\chi^\dagger + \hat{b}_\chi \right)\right]
\hat{d}^\dagger \hat{d},
\end{align}
the acceptor Hamiltonian
\begin{align}
\hat{H}_\text{A}=\sum_{j=1}^N \left\{
\hbar\omega_\text{A}+\sum_\xi g_\xi^\text{A} 
\left[ \left(\hat{b}_\xi^j\right)^\dagger + \hat{b}_\xi^j \right]\right\}
\hat{a}_j^\dagger \hat{a}_j,
\end{align}
the donor-acceptor coupling
\begin{align}
\hat{H}_\text{DA}=-\sum_{j=1}^N
\hbar V_\text{DA}\left(\hat{a}_j^\dagger \hat{d}+\hat{d}^\dagger\hat{a}_j \right),
\label{eq: donor-acceptor coupling}
\end{align}
and the environment Hamiltonian
\begin{align}
\hat{H}_\text{e}=\sum_\chi \hbar\omega_\chi \hat{b}_\chi^\dagger \hat{b}_\chi
+\sum_{j=1}^N \sum_\xi \hbar\omega_\xi \left(\hat{b}_\xi^j\right)^\dagger \hat{b}_\xi^j.
\end{align}
Here, $\hbar\omega_\text{D}$ ($\hbar\omega_\text{A}$) denotes the Franck--Condon excitation energy of the donor (acceptor) molecule, $\hat{d}$ and $\hat{a}_j$ represent the annihilation operators of an electronic excitation at the donor and the $j$th acceptor site, respectively, and $V_\text{DA}$ denotes the strength of the donor--acceptor coupling. 
The frequency, annihilation operator, and coupling strength of the normal mode $\chi$ of the environment to the electronic excitation of the donor are denoted by $\omega_\chi$, $\hat{b}_\chi$, and $g_\chi^\text{D}$, respectively, whereas those of the normal mode $\xi$ of the environment associated with the $j$th acceptor are denoted by $\omega_\xi$, $\hat{b}_\xi^j$ and $g_\xi^\text{A}$, respectively.

The Hamiltonian of the cavity is given by $\hat{H}_\text{c}=\hbar\omega_\text{c}\hat{c}^\dagger\hat{c}$, where $\omega_\text{c}$ and $\hat{c}$ denote the frequency and annihilation operator of a cavity photon, respectively.
The interaction between the cavity and acceptors through dipolar coupling is expressed by the Hamiltonian as follows:
\begin{align}
\hat{H}_\text{I}=-\sum_{j=1}^N \hbar\Omega_\text{R} 
\left(\hat{a}_j^\dagger \hat{c}+\hat{c}^\dagger \hat{a}_j\right),
\end{align}
where $\Omega_\text{R}$ represents the single-molecule Rabi frequency characterizing the interaction strength.
The rotating-wave approximation was used because we did not consider the ultrastrong coupling regime~\cite{Phuc20}.
We considered the case of cavity resonance, namely $\omega_\text{c}=\omega_\text{A}$.
Therefore, the low-energy molecular-exciton polariton, namely the lower polariton, is a superposition of a cavity photon and electronic excitations of all acceptor molecules, and its annihilation operator is expressed as follows:
\begin{align}
\hat{p}=\frac{\hat{c}}{\sqrt{2}}+\frac{1}{\sqrt{2N}}\sum_{j=1}^N \hat{a}_j.
\label{eq: lower polariton state}
\end{align}

In this study the group of acceptors is considered to be prepared in a BEC state with a total number $N_\text{p}$ of molecular polaritons, and the donor is initially in its electronic excited state. 
The initial state of the system is approximately given by
\begin{align}
|\psi_\text{i}\rangle=\frac{1}{\sqrt{N_\text{p}!}}
\left(\hat{p}^\dagger\right)^{N_\text{p}} \hat{d}^\dagger |0\rangle,
\label{eq: molecular polariton BEC state}
\end{align}
where $|0\rangle$ denotes the full ground state of the total system, in which all the molecules are in their electronic ground states.
The $1/\sqrt{N_\text{p}!}$ factor is introduced for state normalization.
Note that in experimentally realized molecular polariton BECs, it is typically that $N_\text{p}/N\ll 1$ such that the inter-exciton interactions can be neglected at the lowest-order approximation as the system is in the weakly interacting BEC regime, as opposed to the Mott-insulator regime where the interactions between excitons become dominant.

\textit{Perturbative limit}--In weak-donor-acceptor-coupling and high-temperature regimes, and if the molecule--cavity coupling is sufficiently strong such that the effects of energy fluctuations on the polaritonic structure can be ignored, the Fermi's golden rule can be used to determine the analytic expression for the EET rate.
The rate of transfer from the initial state $|\psi_\text{i}\rangle$ to the final state $|\psi_\text{f}\rangle=(1/\sqrt{(N_\text{p}+1)!})\left(\hat{p}^\dagger\right)^{N_\text{p}+1} |0\rangle$ is proportional to $|\langle\psi_\text{f}|\hat{H}_\text{DA}|\psi_\text{i}\rangle|^2$. 
We assumed that the Rabi frequency is sufficiently large such that the lower polariton state is energetically separated from the upper polariton and the group of dark states, to which excitation energy transfers can be ignored owing to the off-resonance condition.
From Eq.~\eqref{eq: donor-acceptor coupling} for $\hat{H}_\text{DA}$, the EET rate was determined to be (see Appendix~\ref{appdx: Excitation-energy transfer to a molecular-exciton-polariton condensate} for details)
\begin{align}
k_{\text{D}\to\text{A}}^\text{BEC}\propto (N_\text{p}+1)N\hbar^2 V_\text{DA}^2.
\label{eq: EET rate}
\end{align}
For comparison, in the absence of the molecular polariton BEC, the rate of EET from an exciton donor to a group of $N$ exciton acceptors is given by $k_{\text{D}\to\text{A}}^\text{no-BEC}\propto N\hbar^2 V_\text{DA}^2$. 
It is clear that a Bose enhancement of EET with an increasing factor of $N_\text{p}+1$ can be observed in the system with a molecular-polariton BEC. 

\textit{Numerical simulation}--Subsequently, we numerically investigate the EET dynamics under a more general condition that goes beyond the perturbative and Markovian limits. Environmental dynamics are characterized by their correlation functions
\begin{align}
C(t)=\int_0^\infty \text{d}\omega\, J(\omega) 
\left[ \coth \left( \frac{\beta\omega}{2}\right) \cos\omega t- i\sin\omega t\right],
\end{align}  
where $\beta=1/(k_\text{B}T)$ ($k_\text{B}$ denotes the Boltzmann constant), and $J(\omega)=\sum_\xi g_\xi^2 \delta(\omega-\omega_\xi)$ represents the environmental spectral density. 
The Drude--Lorentz spectral density $J(\omega)=2\lambda\tau\omega/(\tau^2\omega^2+1)$ is considered in the high-temperature limit $k_\text{B}T\tau/\hbar\gg 1$, for which the correlation function $C(t)$ becomes exponential.
Here, $\lambda=\sum_\xi g_\xi^2/(\hbar\omega_\xi)$ is the reorganization energy (per molecule), and $\tau$ is the relaxation time of the environment.
Under this condition, the dynamics of the reduced molecular system are obtained by integrating out the environmental degrees of freedom, yielding the HEOM for the system's density operator $\hat{\rho}_{\mathbf{n}=\mathbf{0}}$ and a set of auxiliary operators $\hat{\rho}_{\mathbf{n}\not=\mathbf{0}}$ with labels $\mathbf{n}=(n_1, \cdots, n_{N+1})$ obtained from a set of non-negative integers as follows:
\begin{align}
\frac{\text{d}\hat{\rho}_\mathbf{n}}{\text{d}t}=&
-\left(\frac{i}{\hbar}\hat{H}_\text{s}^\times +\gamma\sum_{j=1}^{N+1}n_j\right) \hat{\rho}_\mathbf{n}(t) \nonumber\\
&+\sum_{j=1}^{N+1}\left[ i \hat{V}_j^\times \hat{\rho}_{\mathbf{n}_j^+}(t)
+n_j\hat{\Theta}_j \hat{\rho}_{\mathbf{n}_j^-}(t)\right],
\label{eq: HEOM}
\end{align}
where $\gamma=\tau^{-1}$, $\hat{V}_j=\hat{a}_j^\dagger\hat{a}_j$ ($j=1, \cdots, N$) and $\hat{V}_{N+1}=\hat{d}^\dagger\hat{d}$. 
Here, the superoperator $\hat{\Theta}_j$ is given by $\hat{\Theta}_j=\lambda(2i k_\mathrm{B}T \hat{V}_j^\times/\hbar +\gamma \hat{V}_j^\circ)$ with $\hat{V}_j^\times \hat{\rho} \equiv \hat{V}_j\hat{\rho}-\hat{\rho}\hat{V}_j$ and $\hat{V}_j^\circ \hat{\rho} \equiv \hat{V}_j\hat{\rho}+\hat{\rho}\hat{V}_j$.
The following set of parameter values is used in our numerical simulation: $V_\text{DA}=10\;\text{cm}^{-1}$, $T=300\;\text{K}$, $\tau=250\;\text{fs}$, and $\lambda=10\;\text{cm}^{-1}$, which correspond to the high-temperature and strong-energy-fluctuation limits ($V_\text{DA}<\sqrt{\lambda k_\text{B}T}$). 
The donor--acceptor energy difference $\Delta\omega_\text{DA}=\omega_\text{A}-\omega_\text{D}$ is set to be equal to the collective Rabi frequency $\Omega_\text{R}\sqrt{N}=20\;\text{meV}$ such that the donor has equal energy level as the lower polariton state. 
The light--matter interaction is strong in the sense that the collective Rabi frequency is large compared to the energy fluctuation amplitude characterized by $\sqrt{\lambda k_\text{B}T}$.
The effect of cavity loss is not considered explicitly, as we assume that a balance between the processes of cavity pumping and photon loss results in a steady state of the polariton BEC.
The effects of radiative and nonradiative population decay of molecular excitations are also ignored, as their timescales are typically much longer than the timescale of EET~\cite{MayKuhn-book}.
For the initial state, we consider the case in which the donor is vertically excited under the Condon approximation while the group of acceptors is prepared in the molecular-exciton-polariton BEC state, as represented by Eq.~\eqref{eq: molecular polariton BEC state}. 
Because of the effect of polaron decoupling~\cite{Spano15, Herrera16, Phuc19, Takahashi20}, the equilibrium position of the molecular polariton's potential energy surface (PES) is almost equal to that of the acceptors' ground-state PES if the number of acceptors is sufficiently large.
Consequently, the initial matrix elements of the auxiliary operators $\hat{\rho}_{\mathbf{n}\not=\mathbf{0}}$ in the HEOM [Eq.~\eqref{eq: HEOM}] are zero to a good approximation.
Owing to the exponential growth of the number of auxiliary operators required in the HEOM with an increasing number of molecules, the numerical simulation of EET dynamics was only performed for a small-sized system with $N=4$.

\begin{figure}[tbp] 
  \centering
  \includegraphics[keepaspectratio]{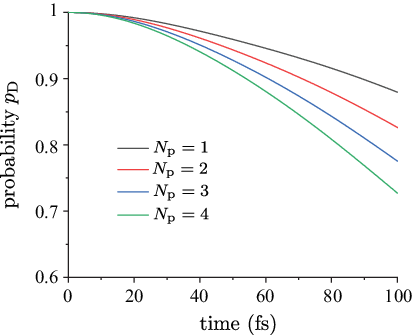}
  \caption{EET dynamics for the different numbers of molecular polaritons in the BEC ($1\leq N_\text{p}\leq 4$). The probability $p_\text{D}$ for finding the donor in its electronic excited state is plotted as a function of time. The group of acceptors is initially prepared in the molecular-exciton-polariton BEC state, as represented by Eq.~\eqref{eq: molecular polariton BEC state}. The parameters of the molecular system and environment are provided in the text.}
  \label{fig: compare EET dynamics for different Np}
\end{figure}

The EET dynamics for different numbers of molecular polaritons in the BEC ($1\leq N_\text{p}\leq 4$) are shown in Fig.~\ref{fig: compare EET dynamics for different Np}. 
The time-dependent probability $p_\text{D}(t)=\text{Tr}\{\hat{d}^\dagger\hat{d}\hat{\rho}_{\mathbf{n}=\mathbf{0}}(t)\}$ to find the donor in its electronic excited state was evaluated up to $t=100\;\text{fs}$.
To focus on the EET in the direction from the donor to the acceptors, we assumed in the simulation that the number of excitations in the donor cannot exceed one because of, for instance, a strong interaction between excitons at the donor site.
Under this assumption, the back flow of excitation energy from acceptors to the donor at short time when the probability $p_\text{D}$ remains close to unity is negligibly small.
It can be observed from Fig.~\ref{fig: compare EET dynamics for different Np} that the EET rate increases with $N_\text{p}$, as expected for a Bose enhancement.
Note that the ratio of $N_\text{p}/N$ in experimentally achieved molecular polariton BECs is usually much smaller than unity.
However, as shown by Eq.~\eqref{eq: EET rate}, the effect of Bose enhancement is determined by the number $N_\text{p}$ of polaritons in the condensate rather than by the ratio of $N_\text{p}/N$.

\begin{figure}[tbp] 
  \centering
  \includegraphics[keepaspectratio]{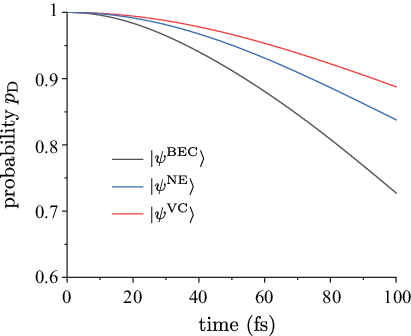}
  \caption{EET dynamics for three different initial states: the molecular-exciton-polariton BEC state $|\psi^\text{BEC}\rangle$, a normal excited state in the group of acceptors $|\psi^\text{NE}\rangle$, and the vacuum state for electronic excitations of acceptors $|\psi^\text{VC}\rangle$, that is, all acceptors are in their electronic ground states. The explicit expressions for these states are provided in the text. As shown in Fig.~\ref{fig: compare EET dynamics for different Np}, the probability $p_\text{D}$ of finding the donor in its electronic excited state is plotted as a function of time. The average number of molecular excitations is set to be the same for $|\psi^\text{BEC}\rangle$ and $|\psi^\text{NE}\rangle$ (equal to two).}
  \label{fig: comparison of EET dynamics for three different initial states}
\end{figure}

Figure~\ref{fig: comparison of EET dynamics for three different initial states} compares the EET dynamics for three different initial states: the molecular-exciton-polariton BEC state $|\psi^\text{BEC}\rangle$ represented by Eq.~\eqref{eq: molecular polariton BEC state} with $N_\text{p}=4$, a normal excited state in the group of acceptors $|\psi^\text{NE}\rangle=\hat{d}^\dagger \prod_{j=1}^{N_\text{e}}\hat{a}_j^\dagger |0\rangle$ with $N_\text{e}=2$, and the vacuum state for electronic excitations of acceptors $|\psi^\text{VC}\rangle=\hat{d}^\dagger|0\rangle$, that is, all acceptors are in their electronic ground states.
Because of the hybridization of light and matter degrees of freedom in molecular polaritons, the average number of molecular excitations is the same for $|\psi^\text{BEC}\rangle$ and $|\psi^\text{NE}\rangle$ (equal to two).
The EET dynamics for the normal excited and vacuum states were calculated under the resonance condition $\Delta\omega_\text{DA}=0$ and zero Rabi coupling $\Omega_\text{R}=0$.
It is clear from Fig.~\ref{fig: comparison of EET dynamics for three different initial states} that the maximum rate of EET is obtained for the molecular-exciton-polariton BEC state, in comparison with the vacuum state and the normal excited state with an equal average number of molecular excitations.
 
\begin{figure}[tbp] 
  \centering
  \includegraphics[keepaspectratio]{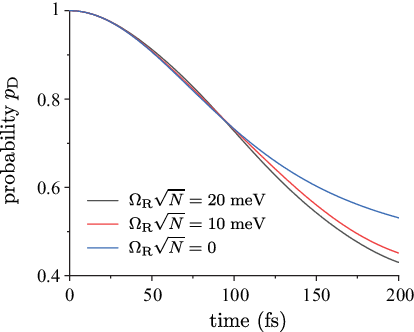}
  \caption{EET dynamics for varying molecule--cavity coupling strengths. The time-dependent probability $p_\text{D}$ for finding the donor in its electronic excited state is plotted for three values of the collective Rabi frequency: $\Omega_\text{R}\sqrt{N}=20\;\text{meV}$, $10\;\text{meV}$, and 0. In all cases, the group of acceptors is initially prepared in the molecular-exciton-polariton BEC state with $N_\text{p}=4$. The other parameters of the system and environment are similar to those in Fig.~\ref{fig: compare EET dynamics for different Np}.}
  \label{fig: Rabi-frequency dependence}
\end{figure}

Finally, Fig.~\ref{fig: Rabi-frequency dependence} shows the time evolution of the exciton population $p_\text{D}(t)$ ($0<t<200\;\text{fs}$) at the donor for varying values of the collective Rabi frequency $\Omega_\text{R}\sqrt{N}$ when the group of acceptors is initially prepared in the molecular-exciton-polariton BEC state with $N_\text{p}=4$.
It can be observed that although the EET dynamics after a short time ($t<100\;\text{fs}$) are almost independent of the molecule--cavity coupling strength, after a longer time, the EET rate becomes smaller for a weaker light--matter interaction.
This decrease in the EET rate can be attributed to the effect of decoherence that can be mitigated by a sufficiently strong molecule--cavity coupling through the polaron decoupling effect~\cite{Phuc21}.
Moreover, the preparation of a molecular-polariton condensate also requires a strong light--matter interaction.

\textit{Effects of the dark-state manifold}--When a group of $N_\text{A}$ acceptor molecules are strongly coupled to a single mode of an optical cavity, in addition to the lower- and upper-polariton states, there exist a manifold consisting of $(N_\text{A}-1)$ dark states, which are not coupled to the cavity field~\cite{Ribeiro18}.
In the case of a donor with equal coupling strengths to all acceptors considered above, it can be shown that the coupling of the donor to any dark state in the group of acceptors would vanish (see Appendix~\ref{appdx: Excitation-energy transfer to the dark-state manifold} for details). 
This is the case, for example, when the molecules are embedded in a nanocavity~\cite{Chikkaraddy16} as the effective range of the dipole-dipole interaction, through which excitation energy is transferred between molecules, is of the order of 10 nm.

In a more typical experimental setup using a microcavity whose size is much larger than the effective range of the dipole-dipole interaction, if the number of donors is small and/or the EETs from different donors are independent, the energy transfer to the acceptor's lower-polariton state can be overwhelmed by that to the acceptor's huge manifold of dark states.
This undesirable effect of dark states has been predicted theoretically for the singlet fission~\cite{Martinez18} and observed experimentally in the intersystem crossing between singlet and triplet excited states~\cite{Eizner19}.

To suppress the EET to the dark-state manifold, we propose a setup in which a group of donors are uniformly distributed in the system of acceptors.
Moreover, in addition to the coupling of all acceptors to a single cavity mode, all the donors are coupled to another cavity mode, which might come from a different optical cavity for a full control.
As a result, another set of polariton and dark states are formed in the donor system, and the EET from the donor's lower-polariton state to the acceptor's polariton BEC is considered in the same way as in the above case of a single donor.
It can be shown that by utilizing the permutation-symmetry property of the dark states, the EET from the donor's polariton state to the acceptor's dark-state manifold would be suppressed and the EET to the acceptor's polariton BEC would dominate (see Appendix~\ref{appdx: Excitation-energy transfer to the dark-state manifold} for details).
Under this condition, the effect of Bose enhancement should be observed.

\textit{Discussion}--It should also be noted that the Bose enhancement cannot be observed if the donor and acceptor parts in the current setup are exchanged, i.e., there would be no enhancement in the EET from an exciton-polariton BEC of a group of donors coupled to a common acceptor.
Here, the transfer rate depends only on the number of excitations in the donor group, no matter if they are independent or if they form a condensate.
Consequently, the Bose enhancement can be used, for example, in the transport from an energy source connected to donors to an energy sink connected to acceptors in the presence of an acceptor's exciton-polariton BEC. 
In this setup, a steady state of energy transport would be established after a transient stage.
The net current of energy is given by the difference between the currents in the forward and backward directions.
The intensity of the backward current, in general, depends on both the intensity of the continuous energy flow from the source to the donors and the inter-exciton interactions at the donor sites. 
However, since only the forward but not the backward current is Bose enhanced, the effect of Bose enhancement can be observed by measuring the net current of energy flow.

Unlike the enhancement of reaction rate in superreaction~\cite{Phuc21} and similar phenomena~\cite{Strek77, Scholes02, Lloyd10, Kassal13, Duque15}, which is based on the constructive interference of different pathways, the underlying mechanism of the Bose-enhanced EET in the presence of a molecular-exciton-polariton BEC is similar to that of the stimulated emission of light in which photons are bosonic particles.
In addition to a polariton BEC, the Bose enhancement of EET can also be observed to some extent in a fully symmetrized state of multi excitons~\cite{Lloyd10}.

\textit{Conclusion}--In conclusion, we investigated the EET in a molecular system with an exciton donor coupled to a group of exciton acceptors that are further strongly coupled to a single mode of an optical cavity. 
When the acceptors were prepared in the molecular-exciton-polariton BEC state, a Bose enhancement of EET was observed, which manifested itself as an increase in the EET rate with an increase in the number of molecular polaritons in the BEC.
The Bose enhancement was also justified by the fact that a greater rate of EET was observed if the acceptors were initially prepared in the molecular-exciton-polariton BEC state than if they were either in their electronic ground states or in a normal excited state with an equal average number of molecular excitations.
Finally, a permutation-symmetry-based approach to suppress the EET to the huge manifold of dark states in the acceptor group was proposed to facilitate the Bose-enhanced EET to the polariton BEC.
As exemplified by the Bose-enhanced EET, room-temperature molecular-polariton condensates are a promising platform for utilizing quantum statistics to manipulate the physical and chemical properties of molecular systems.

\begin{acknowledgements}
This work was supported by JSPS KAKENHI Grant Number 19K14638.
The computations were performed using Research Center for Computational Science, Okazaki, Japan.
\end{acknowledgements}

\section*{Data availability}
The data that support the findings of this study are available from the corresponding author upon reasonable request.

\appendix
\section{EET to a molecular-exciton-polariton condensate}
\label{appdx: Excitation-energy transfer to a molecular-exciton-polariton condensate}
The rate of transfer from the initial state $|\psi_\text{i}\rangle$ given by Eq.~\eqref{eq: molecular polariton BEC state} to the final state
\begin{align}
|\psi_\text{f}\rangle=\frac{1}{\sqrt{(N_\text{p}+1)!}}
\left(\hat{p}^\dagger\right)^{N_\text{p}+1} |0\rangle,
\end{align}
is proportional to $|\langle\psi_\text{f}|\hat{H}_\text{DA}|\psi_\text{i}\rangle|^2$.
The lower-molecular-polariton annihilation operator given by Eq.~\eqref{eq: lower polariton state} can be expressed as
\begin{align}
\hat{p}=\frac{\hat{c}+\hat{A}}{\sqrt{2}},
\label{eq: lower polariton state 2}
\end{align}
where $\hat{A}=(1/\sqrt{N})\sum_{j=1}^N\hat{a}_j$ is the collective annihilation operator of excitation in the group of acceptors.
The donor-acceptor coupling Hamiltonian is given by
\begin{align}
\hat{H}_\text{DA}=&-\hbar V_\text{DA}\sqrt{N}\left(\hat{A}^\dagger \hat{d}+\hat{d}^\dagger\hat{A}\right),
\label{eq: donor-acceptor coupling}
\end{align}

In the following, we will evaluate the matrix element $\langle\psi_\text{f}|\hat{H}_\text{DA}|\psi_\text{i}\rangle$.
It can be seen that since there is no electronic excitation at the donor in the final state, the second term $\hat{d}^\dagger\hat{A}$ in the Hamiltonian $\hat{H}_\text{DA}$ gives a vanishing matrix element $\langle\psi_\text{f}|\hat{d}^\dagger\hat{A}|\psi_\text{i}\rangle$.
For the first term $\hat{A}^\dagger \hat{d}$ in the Hamiltonian $\hat{H}_\text{DA}$, we can use $\hat{A}=\sqrt{2}\hat{p}-\hat{c}$ to get
\begin{align}
\langle\psi_\text{f}|\hat{H}_\text{DA}|\psi_\text{i}\rangle=&
-\hbar V_\text{DA}\sqrt{N}
\Bigg[ \sqrt{2}\langle (N_\text{p}+1)_\text{p}|\hat{p}^\dagger|(N_\text{p})_\text{p}\rangle \nonumber\\
&-\frac{1}{\sqrt{(N_\text{p}+1)!N_\text{p}!}}\langle 0|\hat{p}^{N_\text{p}+1}\hat{c}^\dagger \left(\hat{p}^\dagger\right)^{N_\text{p}}|0\rangle \Bigg],
\label{eq: matrix element}
\end{align}
where the polariton Fock state is defined by $|(n)_\text{p}\rangle=(1/\sqrt{n!})\left(\hat{p}^\dagger\right)^n|0\rangle$.
The first term in Eq.~\eqref{eq: matrix element} yields 
\begin{align}
\langle (N_\text{p}+1)_\text{p}|\hat{p}^\dagger|(N_\text{p})_\text{p}\rangle=\sqrt{N_\text{p}+1}.
\end{align}
The second term can be evaluated by using $\hat{p}=(\hat{c}+\hat{A})/\sqrt{2}$ and making the following binary expansions:
\begin{align}
\hat{c}^\dagger (\hat{p}^\dagger)^{N_\text{p}}|0\rangle=&
\left(\frac{1}{\sqrt{2}}\right)^{N_\text{p}}\sum_{k=0}^{N_\text{p}} 
\frac{N_\text{p}!\sqrt{(k+1)!}}{k!\sqrt{(N_\text{p}-k)!}} \nonumber\\
&\times |(k+1)_\text{c}, (N_\text{p}-k)_\text{a}\rangle
\end{align}
and 
\begin{align}
\left(\hat{p}^\dagger \right)^{N_\text{p}+1}|0\rangle=&
\left(\frac{1}{\sqrt{2}}\right)^{N_\text{p}+1}\sum_{j=0}^{N_\text{p}+1}
\frac{(N_\text{p}+1)!}{\sqrt{j!}\sqrt{(N_\text{p}+1-j)!}} \nonumber\\
&\times |(j)_\text{c}, (N_\text{p}+1-j)_\text{a}\rangle.
\end{align}
Here, the Fock states in the Hilbert space of cavity photons and acceptors' electronic excitations are defined by $|(n)_\text{c}, (m)_\text{a}\rangle=(1/\sqrt{n!m!})\left(\hat{c}^\dagger\right)^n \left(\hat{A}^\dagger\right)^m |0\rangle$.
It can be seen that only terms with $j=k+1$ give nonzero contributions to the matrix element $\langle\psi_\text{f}|\hat{H}_\text{DA}|\psi_\text{i}\rangle$, yielding
\begin{align}
&\frac{1}{\sqrt{(N_\text{p}+1)!N_\text{p}!}}\langle 0|\hat{p}^{N_\text{p}+1}\hat{c}^\dagger \left(\hat{p}^\dagger\right)^{N_\text{p}}|0\rangle \nonumber\\
=&\left(\frac{1}{\sqrt{2}}\right)^{2N_\text{p}+1} \sqrt{(N_\text{p}+1)!N_\text{p}!}
\sum_{k=0}^{N_\text{p}} \frac{1}{k!(N_\text{p}-k)!} \nonumber\\
=&\sqrt{\frac{N_\text{p}+1}{2}}.
\end{align}
Here, we used the identity $\sum_{k=0}^N 1/[k!(N-k)!]=2^N/N!$.

From the above results, we obtain the matrix element 
\begin{align}
\langle\psi_\text{f}|\hat{H}_\text{DA}|\psi_\text{i}\rangle=
-\hbar V_\text{DA}\sqrt{\frac{N(N_\text{p}+1)}{2}},
\label{eq: matrix element 2}
\end{align}
and in turn, the transfer rate $k_{\text{D}\to\text{A}}^\text{BEC}$, which is proportional to
\begin{align}
|\langle\psi_\text{f}|\hat{H}_\text{DA}|\psi_\text{i}\rangle|^2=
\frac{1}{2}\hbar^2 V_\text{DA}^2 N(N_\text{p}+1).
\label{eq: EET rate}
\end{align}

\section{Excitation-energy transfer to the dark-state manifold}
\label{appdx: Excitation-energy transfer to the dark-state manifold}
If the molecule-cavity coupling strengths are equal for all molecules, the annihilation operators for the set of dark states $|\text{D}_l^\text{A}\rangle$ ($l=1,\cdots, N_\text{A}-1$) are given in the second quantization by
\begin{align}
\hat{D}_l^\text{A}=\frac{1}{\sqrt{N_\text{A}}}\sum_{k=1}^{N_\text{A}} e^{i2\pi kl/N_\text{A}}\hat{a}_k.
\label{eq: dark states}
\end{align}
With the donor-acceptor coupling Hamiltonian $\hat{H}_\text{DA}$ given by Eq.~\eqref{eq: donor-acceptor coupling}, it is straightforward to show that the coupling of the donor to any dark state would vanish:
\begin{align}
\langle \text{D}_l^\text{A}|\hat{H}_\text{DA}|\text{d}\rangle
\propto \sum_{k=1}^{N_\text{A}} e^{-i2\pi kl/N_\text{A}}
=0,
\end{align}
where $|\text{d}\rangle=\hat{d}^\dagger|0\rangle$ and $|\text{D}_l^\text{A}\rangle=\left(\hat{D}_l^\text{A}\right)^\dagger|0\rangle$.
This is the case, for example, when the molecules should be embedded in a nanocavity~\cite{Chikkaraddy16} as the effective range of the dipole-dipole interaction, by which excitation energy is transferred between molecules, is of the order of 10 nm.

In a more typical experimental setup with a microcavity whose size is much larger than the effective range of the dipole-dipole interaction, if the number of donors is small and/or the EETs from different donors are independent, the energy transfer to the lower-polariton state can be overwhelmed by that to the manifold consisting of a huge number of dark states. 
To suppress the EET to the dark-state manifold, we propose a setup in which a group of donors are uniformly distributed in the system of acceptors.
Moreover, in addition to the coupling of all acceptors to a single cavity mode, all the donors are coupled to another cavity mode, which might come from a different optical cavity for a full control.
As a result, another set of polariton and dark states are formed in the donor system.
The EET from the donor's lower-polariton state to the acceptor's polariton BEC is considered in the same way as in the above case of a single donor.

We denote the annihilation operators of the group of donors by $\hat{d}_j$ ($j=1, \cdots, N_\text{D}$), where $N_\text{D}$ is the number of donors.
The donor-acceptor interaction Hamiltonian is generalized from Eq.~\eqref{eq: donor-acceptor coupling} to
\begin{align}
\hat{H}_\text{DA}=-\sum_{j=1}^{N_\text{D}} \sum_{k=1}^{N_\text{A}}\hbar V_\text{DA}^{jk}
\left( \hat{a}_k^\dagger \hat{d}_j + \hat{d}_j^\dagger \hat{a}_k\right),
\label{eq: generalized donor-acceptor coupling}
\end{align}
where $\hat{d}_j$ and $\hat{a}_k$ are the annihilation operators of electronic excitations at the $j$th donor and $k$th acceptor, respectively, and their coupling is strength is denoted by $V_\text{DA}^{jk}$.
The annihilation operator of the donor's lower-polariton state is given by
\begin{align}
\hat{p}_\text{D}=\frac{\hat{c}_\text{D}}{\sqrt{2}}
+\frac{1}{\sqrt{2N_\text{D}}}\sum_{j=1}^{N_\text{D}}\hat{d}_j,
\label{eq: donor's polariton state}
\end{align}
where the operator $\hat{c}_\text{D}$ annihilates a photon in the cavity mode coupled to the donors.

For the $k$th acceptor, its coupling strength to the donor's polariton state is given by $V_\text{DA}^k=(1/\sqrt{2N_\text{D}})\sum_{j=1}^{N_\text{D}} V_\text{DA}^{jk}$, which in general varies from one acceptor to another.
However, due to the uniformity of the system, the average value $\bar{V}_\text{DA}$ of this coupling strength should be spatially independent.
We divide the coupling strength $V_\text{DA}^k$ into the average value $\bar{V}_\text{DA}$ and its fluctuation $\delta V_\text{DA}^k$, whose average value vanishes $\langle \delta V_\text{DA}^k\rangle=0$:
\begin{align}
V_\text{DA}^k=\bar{V}_\text{DA}+\delta V_\text{DA}^k.
\end{align}
From Eqs.~\eqref{eq: dark states},~\eqref{eq: generalized donor-acceptor coupling}, and~\eqref{eq: donor's polariton state}, it is straightforward to show that the averaged couplings of the donor's polariton state to the acceptor's dark states $|\text{D}^\text{A}_l\rangle$ ($l=1,\cdots, N_\text{A}-1$) vanish as a result of the dark states' permutation symmetry:
\begin{align}
\langle \text{D}^\text{A}_l|\hat{\bar{H}}_\text{DA}|\text{p}_\text{D}\rangle=0,
\end{align}
where $|\text{p}_\text{D}\rangle=\hat{p}_\text{D}^\dagger|0\rangle$.

On the other hand, it can be shown that the contribution of the fluctuation $\delta V_\text{DA}^k$ of the donor-acceptor coupling strength to the EET from the donor's polariton state to the acceptor's manifold of dark states cannot overwhelm the EET to the acceptor's polariton state.
Indeed, assuming independent behaviors of the acceptors, the variance of the sum $\delta V_\text{DA}=\sum_{k=1}^{N_\text{A}} \delta V_\text{DA}^k$ scales as $N_\text{A}^{1/2}$ with the number of acceptors.
Due to the factor of $N_\text{A}^{-1/2}$ in Eq.~\eqref{eq: dark states} for the dark states, the coupling strength of the donor's polariton state to one dark state of the acceptors as a result of fluctuation in coupling strength would scale as $N_\text{A}^0$.
Taking into account the fact that there are a total of $(N_\text{A}-1)$ dark states in the acceptor's manifold, the total rate of EET from the donor's polariton state to the acceptor's dark-state manifold would scale as $N_\text{A}^1$.
This scaling is the same as for the rate of EET to the acceptor's polariton state since the coupling strength of the polariton state is enhanced by a factor of $N_\text{A}^{1/2}$, as shown, for example, by Eqs.~\eqref{eq: matrix element} and~\eqref{eq: EET rate}.
The effect of Bose enhancement and the large separation in energy between the acceptor's lower-polariton and dark states would then strongly suppress the EET from the donor's polariton state to the acceptor's dark-state manifold to favor the EET to the acceptor's polariton BEC.


\end{document}